\documentclass[a4paper]{report}
\usepackage[utf8]{inputenc}
\usepackage[T1]{fontenc}
\usepackage{RJournal}
\usepackage{amsmath,amssymb,array}
\usepackage{booktabs}

\begin{document}

\sectionhead{Contributed research article}
\volume{XX}
\volnumber{YY}
\year{20ZZ}
\month{AAAA}

\begin{article}
\title{\pkg{quantreg.nonpar}: An R Package for Performing Nonparametric Series Quantile Regression}
\author{by Michael Lipsitz, Alexandre Belloni, Victor Chernozhukov, and Iv\'{a}n
Fern\'{a}ndez-Val}

\maketitle

\abstract{The R package \pkg{quantreg.nonpar} implements nonparametric 
quantile regression methods to estimate and make inference on partially linear quantile models.  \pkg{quantreg.nonpar}
obtains point estimates  of the conditional quantile
function and its derivatives based on 
series approximations to the nonparametric part of the model. It also provides pointwise and uniform confidence
intervals over a region of covariate values and/or quantile indices for the same functions using analytical and resampling methods.
This paper serves as an introduction to the package and displays
basic functionality of the functions contained within.
}

\section{Introduction: nonparametric series quantile regression}

Let $Y$ be an outcome variable of interest, and $X$ a vector of observable covariates. The covariate vector is  partitioned  as $X = (W,V),$ where $W$ is the key covariate or treatment, and $V$ is a possibly high dimensional vector with the rest of the covariates that usually play the role of control variables. We can model the $\tau$-quantile of $Y$ conditional on $X=x$ using the partially linear quantile model
$$
Q_{Y \mid X}\left(\tau \mid x\right)=g\left(\tau,w\right)+v'\gamma\left(\tau\right), \ \ \tau \in [0,1].
$$
\cite{bcf11} developed the nonparametric series quantile regression (QR) approximation
$$
Q_{Y \mid X}\left(\tau \mid x\right) \approx Z\left(x\right)'\beta\left(\tau\right), \ \ \beta\left(\tau\right) = \left(\alpha\left(\tau\right)',\gamma\left(\tau\right)'\right)', \ \ Z\left(x\right) = \left(Z\left(w\right)',v'\right)',
$$
where the unknown function $g(\tau,w)$ is approximated by a linear combination of series terms   $Z(w)'\alpha(\tau)$. The vector $Z(w)$ includes transformations of $w$ that have good approximation properties such as powers, indicators, trigonometric terms or B-splines. The function $\tau \mapsto \alpha(\tau)$ contains quantile-specific coefficients. The  \CRANpkg{quantreg.nonpar}
package implements estimation and inference method for linear functionals of the conditional quantile function based on the series QR approximation. These functionals include:
\begin{enumerate}
\item Conditional quantile function itself: $(\tau,x) \mapsto Q_{Y \mid X}(\tau \mid x) \approx Z(x)'\beta(\tau).$
\item Partial first and second derivative functions with respect to $w$: $(\tau,x) \mapsto \partial^k Q_{Y \mid X}(\tau \mid x) / \partial w^k = \partial^k g(\tau,w) / \partial w^k \approx  \partial^k Z(w)'\beta(\tau) / \partial w^k ,$ $k \in \{1,2\}.$
\item Average partial first and second derivative functions with respect to $w$: \\ $(\tau) \mapsto \int \partial^k g(\tau,w) / \partial w^k d\mu \approx  \int  \partial^k Z(w)'\beta(\tau) / \partial w^k d\mu,$ $k \in \{1,2\},$ where $\mu$ is a measure for $W$.
\end{enumerate}
Both pointwise or uniform inference over a region of quantile indexes and/or covariate values are implemeted.

The  coefficient vector $\beta(\tau)$ is estimated using the QR estimator of \citet{koenker:1978}.  Let $\left\{ (Y_{i},X_{i}):1\leq i\leq n\right\} $
be a random sample from $(Y,X)$ and let $\hat{\beta}(\tau)$ be the QR estimator of $\beta(\tau)$, i.e., 
\[
\hat{\beta}\left(\tau\right)\in{\displaystyle \arg\min_{\beta \in \mathbb{R}^m}}{\displaystyle \sum_{i=1}^{n}}\rho_{\tau}\left(Y_{i}-Z\left(X_{i}\right)'\beta\right), \ \ \tau\in\mathcal{{T}}\subseteq\left(0,1\right),
\]
where $\rho_{\tau}(z)=(\tau-1\{z<0\})z$ is the check function, $\mathcal{T}$ is a compact set, and $m = \dim \beta(\tau)$. We then construct estimators of the linear functionals of the conditional quantile function by applying the plug-in principle to the series approximations. For example, the series QR quantile estimator of $Q_{Y \mid X}(\tau \mid x)$ is
$$
\hat Q_{Y \mid X}\left(\tau \mid x\right) = Z\left(x\right)'\hat \beta\left(\tau\right).
$$

A challenge to perform inference in this setting is that $m$ should increase with the sample size in order to reduce approximation error.  Accordingly, the empirical series QR coefficient process $\tau \mapsto \sqrt{n}(\widehat{\beta}(\tau)-\beta(\tau))$ has increasing dimension with $n$ and therefore does not have a limit distribution.  \cite{bcf11} dealed with this problem by deriving two couplings or strong approximations to  $\tau \mapsto \sqrt{n}(\widehat{\beta}(\tau)-\beta(\tau))$. A  coupling is a construction of two processes on the same probability space that uniformly close to each other with high probability. In this case,  \cite{bcf11} constructed a pivotal process and a Gaussian process of dimension $m$ that are uniformly close to  $\tau \mapsto \sqrt{n}(\widehat{\beta}(\tau)-\beta(\tau))$. They also provide four methods to estimate the distribution of these coupling processes that can be used to make inference on linear functionals of the conditional quantile function:
\begin{enumerate}
\item Pivotal: analytical method based on the pivotal coupling.
\item Gradient bootstrap: resampling method based on the pivotal coupling.
\item Gaussian: analytical method based on the Gaussian coupling. 
\item Weighted bootstrap: resampling method based on the Gaussian coupling.
\end{enumerate}
The  \pkg{quantreg.nonpar} package implements all these methods.

Additionally, the linear functionals of interest might be naturally monotone in some of their arguments. For example, the conditional quantile function $\tau \mapsto Q_{Y \mid X}(\tau \mid x) $ is increasing, and in the growth chart application of next section the conditional quantile function of height, $(\tau,x) \mapsto Q_{Y \mid X}(\tau \mid x),$ is increasing with respect to both the quantile index, $\tau$, and the treatment age, $w$. The series QR estimates might not satisfy this logical monotonicity restrictions giving rise to so-called quantile crossing problem in the case of $\tau \mapsto \hat Q_{Y \mid X}(\tau \mid x)$. The  \pkg{quantreg.nonpar} package deals with the quantile crossing and other non monotonicity problems in the estimates of the linear functionals by applying the rearrangement method of \cite{cfg09} and \cite{cfg10}. 

\subsection{Related R packages}

Several existing R packages are available to estimate conditional quantile models. 
The package \CRANpkg{quantreg} (\cite{quantreg2016}) includes multiple commands for parametric and nonparametric quantile regression. The command \code{rqss} estimates univariate and bivariate local nonparametric smoothing splines, and the command \code{rearrange} implements the rearrangement method to tackle the quantile crossing problem.  The package \CRANpkg{QuantifQuantile} (\cite{QuantifQuantile2015}) estimates univariate conditional quantile models using a local nonparametric  method called optimal quantization or partitioning. The nonparametric methods implemented in the previous packages are local or kernel-type, whereas our methods are global or series-type.
Finally, the command \code{gcrq} in the package \CRANpkg{quantregGrowth} (\cite{quantregGrowth2013}) implements a univariate B-spline global  nonparametric method with a penalty to deal with the quantile crossing and impose monotonicity with respect to the covariate. 
To our knowledge, no existing R package allows the user to perform uniform nonparametric inference on linear functionals of the conditional quantile function over a region of quantile indexes and/or covariate values, making \pkg{quantreg.nonpar} the first package to do so.

\section{The package \pkg{quantreg.nonpar}}\label{sec:main}

\subsection{Model specification}

We illustrate the functionality of the package with an empirical application based on data from \cite{koenker2011} for childhood malnutrition in India, where
we model the effect of a child's age and other covariates on the child's
height. 
Here, $Y$ is the child's height in centimeters; $W$ is the child's
age in months; 
and $V$ is a vector of 22 controls. These controls include
the mother's body mass index (BMI), the number of months the child
was breastfed, and the mother's age (as well as the square of the
previous three covariates); the mother's years of education and the
father's years of education; dummy variables for the child's sex,
whether the child was a single birth or multiple birth, whether
or not the mother was unemployed, whether the mother's residence is
urban or rural, and whether the mother has each of: electricity, a
radio, a television, a refrigerator, a bicycle, a motorcycle, and
a car; and factor variables for birth order of the child, the mother's
religion and quintiles of wealth.

First, we load the data and construct the variables that will be used in the analysis.
Note that the variable prefixes "c" and "m" refer to "child"
and "mother". For each factor variable (\code{csex, ctwin, \\ cbirthorder,
munemployed, mreligion, mresidence, wealth, electricity, radio, television, \\
refrigerator, bicycle, motorcycle,} and \code{car}), we generate
a variable "\code{facvar}" which is the factor version of the
variable "\code{var}". For each quadratic variable (\code{mbmi, breastfeeding,} and \code{mage}), we generate a variable "\code{varsq}" which is the variable squared. For example:
\begin{example}
data <- india
faccsex <- factor(csex)
mbmisq <- mbmi^2
\end{example}

We also construct the formula to be used for the linear part of
the model, $v'\gamma(\tau)$:

\begin{example}
form.par <- cheight ~ mbmi + mbmisq + breastfeeding + breastfeedingsq + mage + magesq
	+ medu + edupartner + faccsex + facctwin + faccbirthorder + facmunemployed
	+ facmreligion + facmresidence + facwealth + facelectricity + facradio 
	+ factelevision + facrefrigerator + facbicycle + facmotorcycle + faccar
\end{example}

Note that this formula does not contain a term for our variable of
interest $W$; namely, the child's age. Let us now construct the nonparametric
bases that will be used to estimate the effect of $W$, i.e., $g(\tau,w)\approx Z(w)'\alpha(\tau)$.
For our base case, we construct a cubic B-spline
basis with knots at the $\left\{ 0,0.1,0.2,...,0.9,1\right\} $ quantiles
of the observed values of child's age.

\begin{example}
basis.bsp <- create.bspline.basis(breaks = quantile(cage, c(0:10)/10))
\end{example}

Finally, we set the values of some of the other parameters. For the
purposes of this example, we use 500 simulations for the pivotal and
Gaussian methods, and 100 repetitions for the weighted and gradient
bootstrap methods. The set of analyzed quantile indices will be $\left\{ 0.04,0.08,...,0.96\right\} $,
but we will have \code{npqr} print only results for quantile indices
contained in the set $\left\{ 0.2,0.4,0.6,0.8\right\} $. Finally,
we will use $\alpha=0.05$ as the significance level for the confidence intervals (i.e., the confidence level is 0.95).

\begin{example}
B <- 500
B.boot <- 100
taus <- c(1:24)/25
print.taus <- c(1:4)/5
alpha <- 0.05
\end{example}

\subsection{Comparison of the inference processes}\label{sec:compInf}

Initially, we will focus on the average growth rate, i.e., the average first derivative
of the conditional quantile function with respect to child's age 
$$\tau \mapsto \int \partial_w g\left(\tau,w\right) d \mu\left(w\right), \ \ \tau \in \mathcal{T},$$
where $\mu$ is a measure for $W$ and $\mathcal{T}$ is the set of quantile indices of interest specified with \code{taus}. We specify the average first derivative with the options \code{nderivs = 1} and  \code{average = 1}. 
Inference will be performed  uniformly over $\mathcal{T}$, and the standard errors will be computed
unconditionally for the pivotal and Gaussian processes; see Section \nameref{subsec:cis}.

We first construct the four inference processes based on the B-spline basis. By default, \code{npqr} generates output similar to that seen below. In this example, output is suppressed in each call following the first. Instead of invoking a particular process, we may also set \code{process="none"}. In that case, inference will not be performed, and only point estimates will be reported.

\begin{example}
piv.bsp <- npqr(formula = form.par, basis = basis.bsp, var = "cage", taus = taus,
	nderivs = 1, average = 1, print.taus = print.taus, B = B, uniform = T)
gaus.bsp <- update(piv.bsp, process = "gaussian", printOutput = F)
wboot.bsp <- update(gaus.bsp, process = "wbootstrap", B = B.boot)
gboot.bsp <- update(wboot.bsp, process = "gbootstrap")
\end{example}

The output for the pivotal method (which is generated whenever \code{printOutput = T}) is:

\noindent \begin{center}
\includegraphics[width = \textwidth]{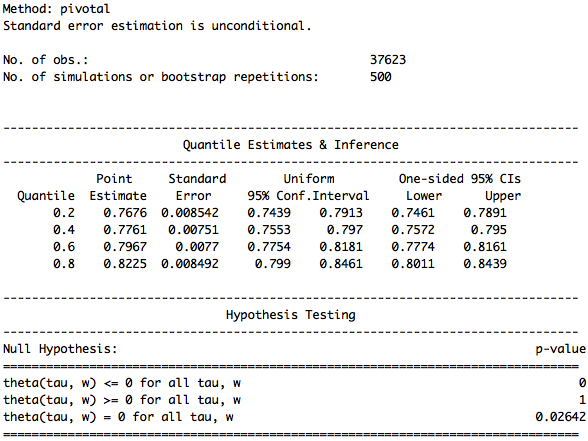}
\par\end{center}

The point estimates represent the average derivative of the conditional quantile function with respect to the variable of interest: the child's age. In other words, each value represents the average rate of growth (in centimeters per month) at each quantile of the height distribution. They are reported, along with their standard errors and respective two-sided and one-sided confidence intervals, at each quantile for which output was requested using \code{print.taus}. The null hypotheses  on which hypothesis testing is performed state that the average growth rate is negative, positive, and equal to zero, respectively,  at all quantiles of the distribution. We reject, at the 5\% level, the null hypotheses that the growth rate is negative and that the growth rate is equal to zero. We can not reject, at the 5\% level, the null hypothesis that the growth rate is positive.

Additionally, the following results are saved in \code{piv.bsp}:
\begin{itemize}
\item \code{piv.bsp\$CI}: a 1 $\times$ \code{length(taus)} $\times$ 2 array: each pair is the lower and
upper bounds of the 95\% confidence interval for the
average derivative of the conditional quantile
function at each quantile index in \code{taus}
\item \code{piv.bsp\$CI.oneSided}: a 1 $\times$ \code{length(taus)} $\times$ 2 array: each pair contains
bounds for two separate one-sided 95\% confidence intervals (a lower
bound and an upper bound, respectively) for the average derivative
of the conditional quantile function at each quantile index in \code{taus}
\item \code{piv.bsp\$point.est}: a 1 $\times$ \code{length(taus)}  matrix: each entry is the point
estimate for the average derivative of the 
conditional quantile function at each quantile index in \code{taus}
\item \code{piv.bsp\$std.error}: a 1 $\times$ \code{length(taus)}  matrix: each entry is the standard
error of the estimator of the average derivative
of the conditional quantile function at each quantile index in \code{taus} (here, unconditional on the sample)
\item \code{piv.bsp\$pvalues}: a three item vector containing the p-values
reported above: the first tests the null hypothesis that the average
derivative is  less than zero everywhere (at each quantile index in \code{taus}); the second tests the null
hypothesis that the average derivative is everywhere greater than
zero; the third tests the null hypothesis that the average derivative
is everywhere equal to zero
\item \code{piv.bsp\$taus}: the input vector \code{taus}, i.e., $\left\{ 0.04,0.08,...,0.96\right\} $
\item \code{piv.bsp\$coefficients}: a list of length \code{length(taus)}: each element of the list contains the estimates of the QR coefficient vector $\beta(\tau)$ at the corresponding
quantile index
\item \code{piv.bsp\$var.unique}: a vector containing all values of the
covariate of interest, $W$, with no repeated values
\item \code{piv.bsp\$load}: the input vector or matrix \code{load}. If \code{load} is not input (as in this case), the output \code{load} is generated based on \code{average} and \code{nderivs}. Here, it is a vector containing the average value of the derivative of the regression equation with respect to the variable of interest, not including the estimated coefficients.
\end{itemize}

Using \code{piv.bsp\$taus}, \code{piv.bsp\$CI}, and \code{piv.bsp\$point.est},
as well as the corresponding objects for the Gaussian, weighted bootstrap,
and gradient bootstrap methods, we construct plots containing
the estimated average quantile derivatives, as well as 95\% uniform confidence
bands over the quantile indices in \code{taus}:

\begin{example}
par(mfrow = c(2, 2))
yrange <- c(.65, .95)
xrange <- c(0, 1)
plot(xrange, yrange, type = "n", xlab = "Quantile Index", 
	ylab = "Average Growth (cm/month)", ylim = yrange)
lines(piv.bsp$taus, piv.bsp$point.est)
lines(piv.bsp$taus, piv.bsp$CI[1, , 1], col = "blue")
lines(piv.bsp$taus, piv.bsp$CI[1, , 2], col = "blue")
title("Pivotal")
plot(xrange, yrange, type = "n", xlab = "Quantile Index", ylab = "", ylim = yrange)
lines(gaus.bsp$taus, gaus.bsp$point.est)
lines(gaus.bsp$taus, gaus.bsp$CI[1, ,1], col="blue")
lines(gaus.bsp$taus, gaus.bsp$CI[1, ,2], col="blue")
title("Gaussian")
plot(xrange, yrange, type = "n", xlab = "Quantile Index", 
	ylab = "Average Growth (cm/month)", ylim = yrange)
lines(wboot.bsp$taus, wboot.bsp$point.est)
lines(wboot.bsp$taus, wboot.bsp$CI[1, , 1], col = "blue")
lines(wboot.bsp$taus, wboot.bsp$CI[1, , 2], col = "blue")
title("Weighted Bootstrap")
plot(xrange, yrange, type = "n", xlab = "Quantile Index", ylab = "", ylim = yrange)
lines(gboot.bsp$taus, gboot.bsp$point.est)
lines(gboot.bsp$taus, gboot.bsp$CI[1, , 1], col = "blue")
lines(gboot.bsp$taus, gboot.bsp$CI[1, , 2], col = "blue")
title("Gradient Bootstrap")
title("Average Growth Rate with 95
\end{example}

\begin{figure}[!hp]
\begin{center}
\includegraphics[width=\textwidth]{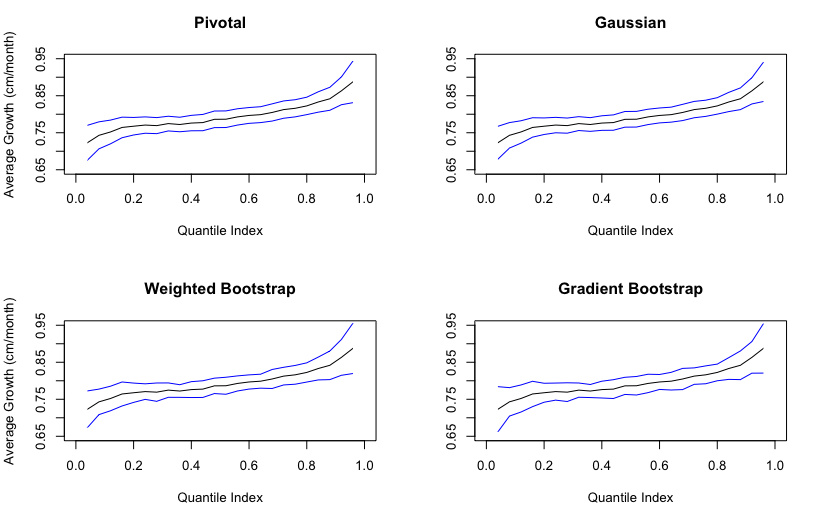}
\caption{\label{fig:growth_rate} Comparison of Inference Methods for Growth Rate: point estimates and 95\% uniform confidence bands for the average derivative of the conditional quantile function of height with respect to age based on B-spline series approximation.}
\end{center}
\end{figure}

As we can see in Figure \ref{fig:growth_rate}, the confidence bands generated are roughly similar.
Note that the point estimates are the same for all the methods.

We can compare the computation times of each of the approximations
using the command \code{Sys.time}. Additionally, we compare
the p-values generated by each of the four inference methods. Note
that computation times may vary widely depending on the machine in
use. However, the relative computation times will be approximately
constant across different machines. The computation times in the
table below were obtained on a computer with two eight-core 2.6 GHz
processors (note: \code{npqr} does not make use of parallel
computing).

\begin{example}
pval.dimnames <- vector("list", 2)
pval.dimnames[[1]] <- c("Pivotal", "Gaussian", "Weighted Bootstrap",
	"Gradient Bootstrap")
pval.dimnames[[2]] <- c("H0: Growth Rate <= 0", "H0: Growth Rate >= 0",
	"H0: Growth Rate = 0", "Computation Minutes")
pvals <- matrix(NA, nrow = 4, ncol = 4, dimnames = pval.dimnames)
pvals[1,] <- c(round(piv.bsp$pvalues, digits = 4), round(piv.time, digits = 0))
pvals[2,] <- c(round(gaus.bsp$pvalues, digits = 4), round(gaus.time, digits = 0))
pvals[3,] <- c(round(wboot.bsp$pvalues, digits = 4), round(wboot.time, digits = 0))
pvals[4,] <- c(round(gboot.bsp$pvalues, digits = 4), round(gboot.time, digits = 0))
print(pvals)
\end{example}

These commands generate the output:

\begin{tabular}{lcccc}
 &
\code{H0: Growth} &
\code{H0: Growth} &
\code{H0: Growth} &
\code{Computation}\tabularnewline
 &
\code{Rate <= 0} &
\code{Rate >= 0} &
\code{Rate = 0} &
\code{Minutes}\tabularnewline
\code{Pivotal} &
\code{0} &
\code{1} &
\code{0.0237} &
\code{0.9}\tabularnewline
\code{Gaussian} &
\code{0} &
\code{1} &
\code{0.0234} &
\code{0.6}\tabularnewline
\code{Weighted Bootstrap} &
\code{0} &
\code{1} &
\code{0.0221} &
\code{30}\tabularnewline
\code{Gradient Bootstrap} &
\code{0} &
\code{1} &
\code{0.0221} &
\code{346}\tabularnewline
\end{tabular}

As expected, we reject at the 5\% level the null hypothesis that the growth rate is
negative and the null hypothesis that the growth rate is equal
to zero in all cases, and we fail to reject the null hypothesis that
the growth rate is positive in all cases. For the one-sided
tests, the relevant null hypothesis is that the average growth rate
is less than or equal to zero (greater than or equal to zero) at all the quantile indices in \code{taus}. For the two-sided
test, the relevant null hypothesis is that the average growth rate
is equal to zero at all the quantile indices in \code{taus}. Additionally, note that the pivotal and Gaussian methods are substantially faster than the two bootstrap methods.

\subsection{Comparison of series bases}\label{sec:compSeries}

Another  option is to take advantage of the variety of
bases available in the \pkg{quantreg.nonpar} package. Here, we consider
three bases: the B-spline basis used in the analysis above, an orthogonal polynomial
basis of degree 12, and a Fourier basis with 9 basis functions and a period of 200 months.
We compare the estimates of the average quantile derivative function generated by
using each of these bases. We construct the orthogonal polynomial basis and the Fourier basis with the commands:

\begin{example}
basis.poly <- poly(cage, degree = 12)
basis.four <- create.fourier.basis(rangeval = range(data$cage), nbasis = 9, 
                     period = 200)
\end{example}

In this section, we focus on the pivotal method for inference.  
We run \code{npqr} for the orthogonal polynomial basis and the Fourier basis, mimicking the analysis run above for the B-spline basis.

\begin{example}
piv.poly <- update(piv.bsp, basis = basis.poly)
piv.four <- update(piv.bsp, basis = basis.four)
\end{example}

\begin{figure}
\begin{center}
\includegraphics[width = \textwidth]{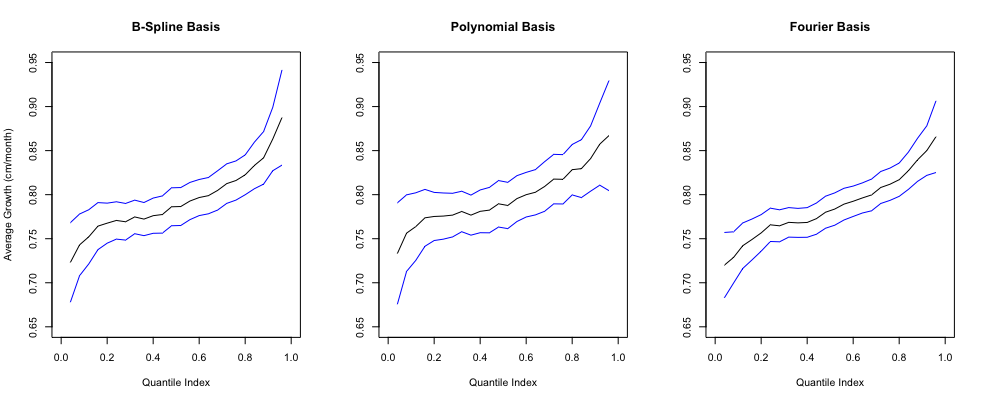}
\end{center}
\caption{\label{fig:growth_rate2} Comparison of Series Bases for Growth Rate: point estimates and 95\% uniform confidence bands for the average derivative of the conditional quantile function of height with respect to age based on B-spline, polynomial, and Fourier series approximations.}
\end{figure}

Similar to Section \nameref{sec:compInf}, we plot the point estimates with their uniform
95\% confidence bands for each basis.
Figure \ref{fig:growth_rate2} shows that, given the parameters of the chosen bases, the type of basis does not have an important impact on the estimation and inference on the growth rate charts. A table containing the p-values associated with the hypothesis tests for each basis are generated by the following code:

\begin{example}
pval2.dimnames <- vector("list", 2)
pval2.dimnames[[1]] <- c("B-spline", "Polynomial", "Fourier")
pval2.dimnames[[2]] <- c("H0: Growth Rate <= 0", "H0: Growth Rate >= 0",
	"H0: Growth Rate = 0")
pvals2 <- matrix(NA, nrow = 3, ncol = 3, dimnames = pval2.dimnames)
pvals2[1,] <- round(piv.bsp$pvalues, digits = 4)
pvals2[2,] <- round(piv.poly$pvalues, digits = 4)
pvals2[3,] <- round(piv.four$pvalues, digits = 4)
print(pvals2)
\end{example}

These commands yield:

\begin{tabular}{cccc}
 &
\code{H0: Growth} &
\code{H0: Growth} &
\code{H0: Growth} \tabularnewline
 &
\code{Rate <= 0} &
\code{Rate >= 0} &
\code{Rate = 0} \tabularnewline
\code{B-Spline} &
\code{0} &
\code{1} &
\code{0.0239}\tabularnewline
\code{Polynomial} &
\code{0} &
\code{1} &
\code{0.0334}\tabularnewline
\code{Fourier} &
\code{0} &
\code{1} &
\code{0.0386}\tabularnewline
\end{tabular}

For all bases, the tests' conclusions are identical: at the 5\% level, we reject the null hypothesis that the average growth rate is negative, fail to reject the null hypothesis that the average growth rate is positive, and reject the null hypothesis that the average growth rate is equal to zero.

\subsection{Confidence intervals and standard errors}\label{subsec:cis}

Now, we illustrate two additional options available to the user. First,
to perform  inference  pointwise over a region of covariate values and/or quantile indices instead of uniformly, and
second, to estimate the standard errors conditional on the values
of the covariate $W$ in the sample. When inference is 
uniform, the test statistic used in construction of the confidence
interval is the maximal t-statistic across all covariate values and quantile indices in the region of interest, whereas  pointwise inference uses the t-statistic at each covariate value and quantile index. When standard
errors are estimated unconditionally, a correction term is used to
account for the fact that the empirical distribution of $W$ is an estimator of the distribution of $W$.
The option to estimate standard errors conditionally or unconditionally
is not available for the bootstrap methods. The inference based on these methods is always unconditional. 

We will use only the pivotal method with a B-spline basis for this
illustration. First, we run \code{npqr} for each combination
of options mentioned above:

\begin{example}
piv.bsp <- npqr(formula = form.par, basis = basis.bsp, var = "cage", taus = taus,
	B = B, nderivs = 1, average = 1, alpha = alpha, process = "pivotal", 
	uniform = T, se = "unconditional", printOutput = F)
piv.bsp.cond <- update(piv.bsp, se = "conditional")
piv.bsp.point <- update(piv.bsp,  uniform = F, se = "unconditional")
piv.bsp.point.cond <- update(piv.bsp,  uniform = F, se = "conditional")
\end{example}

\begin{figure}[h]
 \begin{center}
\includegraphics[width = \textwidth]{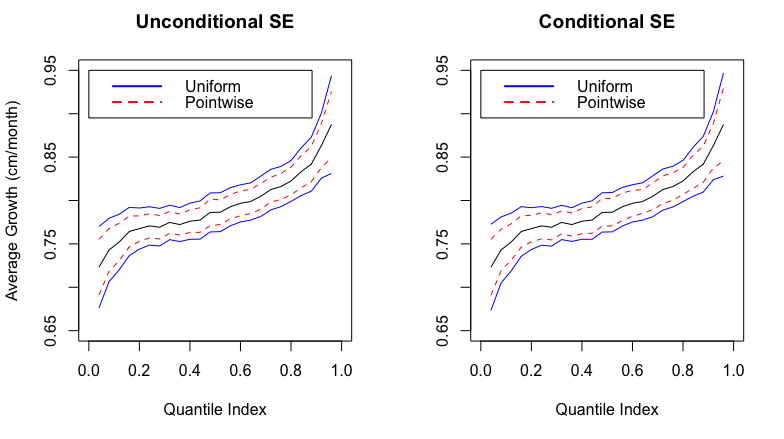}
\caption{\label{fig:growth_rate3} Comparison of Pointwise vs Uniform and Conditional vs Unconditional Inference for Growth Rate: 95\%  uniform and pointwise confidence bands for the average derivative of the conditional quantile function of height with respect to age based on B-spline series approximation. The left panel uses unconditional standard errors in the construction of the bands. The right panel uses conditional standard errors.}
\end{center}
\end{figure}

We obtain Figure \ref{fig:growth_rate3} using the graphing techniques described in Sections \nameref{sec:compInf} and \nameref{sec:compSeries}. As is visible in this figure, usage of conditional standard errors changes the confidence bands only minimally in our example. As expected, the pointwise confidence bands are narrower than the uniform confidence bands.

We can also compare  how much of the differences (or lack thereof) in the confidence bands are driven by differences in the standard errors versus the test statistics. Here, we compare the
estimated standard errors at the median for conditional versus
unconditional inference:

\begin{example}
piv.bsp.med <- npqr(formula = form.par, basis = basis.bsp, var = "cage", taus = 0.5,
	B = B, nderivs = 1, average = 1, alpha = alpha, process = "pivotal", uniform=T,
	se = "unconditional", printOutput = F)
piv.bsp.cond.med <- update(piv.bsp.med, se = "conditional")
stderr.dimnames <- vector("list",2)
stderr.dimnames[[1]] <- c("Unconditional", "Conditional")
stderr.dimnames[[2]] <- c("Standard Error")
stderr <- matrix(NA, nrow = 2, ncol = 1, dimnames = stderr.dimnames)
stderr[1,] <- piv.bsp.med$std.error[1] 
stderr[2,] <- piv.bsp.cond.med$std.error[1]
print(stderr)
\end{example}

These commands yield the output:

\begin{tabular}{cc}
 &
\code{Standard Error}\tabularnewline
\code{Unconditional} &
\code{0.008104}\tabularnewline
\code{Conditional} &
\code{0.007663}\tabularnewline
\end{tabular}

Finally, we compare p-values generated by each of the option choices:

\begin{example}
pval3.dimnames <- vector("list", 2)
pval3.dimnames[[1]] <- c("Uniform, Unconditional", "Uniform, Conditional", 
	"Pointwise, Unconditional", "Pointwise, Conditional")
pval3.dimnames[[2]] <- c("H0: Growth Rate <= 0", "H0: Growth Rate >= 0",
	"H0: Growth Rate = 0")
pvals3 <- matrix(NA, nrow = 4, ncol = 3, dimnames = pval3.dimnames)
pvals3[1,] <- round(piv.bsp$pvalues, digits = 4)
pvals3[2,] <- round(piv.bsp.cond$pvalues, digits = 4)
pvals3[3,] <- round(piv.bsp.point$pvalues, digits = 4)
pvals3[4,] <- round(piv.bsp.point.cond$pvalues, digits = 4)
print(pvals3)
\end{example}

\begin{tabular}{cccc}
 &
\code{H0: Growth} &
\code{H0: Growth} &
\code{H0: Growth}\tabularnewline
 &
\code{Rate <= 0} &
\code{Rate >= 0} &
\code{Rate = 0}\tabularnewline
\code{Uniform, Unconditional} &
\code{0} &
\code{1} &
\code{0.0239}\tabularnewline
\code{Uniform, Conditional} &
\code{0} &
\code{1} &
\code{0.0243}\tabularnewline
\code{Pointwise, Unconditional} &
\code{0} &
\code{1} &
\code{0.0267}\tabularnewline
\code{Pointwise, Conditional} &
\code{0} &
\code{1} &
\code{0.0222}\tabularnewline
\end{tabular}


In this example, where the sample size is large, about 38,000 observations, conditional versus unconditional standard errors
and uniform versus pointwise inference have little impact on the estimated p-values.

\subsection{Estimation and uniform inference on linear functionals}\label{subsec:unif}

Finally, we illustrate how to estimate and make uniform inference on linear functionals of the conditional quantile function over a region of covariate values and quantile indices. These functionals include the  function itself and derivatives  with respect to the covariate of interest. The \pkg{quantreg.nonpar} package is able to perform estimation and inference on the conditional quantile function, its first derivative, and its second derivative over a region of covariate values and/or quantile indices.  
We also illustrate how to report the estimates using three dimensional plots. 

First, we consider the first and second derivatives of the conditional quantile function. In the application they correspond to the growth rate and growth acceleration of height
with respect to age as a function of age (from 0 to 59 months) and the quantile index.
To do so, we use the output of \code{npqr} called \code{var.unique},  which contains a vector with all the distinct values of the covariate of
interest (\code{cage} here). To generate this output, we estimate  the first and second derivatives of the conditional quantile function using a B-spline series approximation over the covariate values in \code{var.unique} and the quantile indices in \code{taus}:

\begin{example}
piv.bsp.firstderiv <- npqr(formula = form.par, basis = basis.bsp, var = "cage",
	taus = taus, nderivs = 1, average = 0, print.taus = print.taus, B = B, 
	process = "none", printOutput = F)
piv.bsp.secondderiv <- update(piv.bsp.firstderiv, nderivs = 2)
\end{example}

Next, we generate vectors containing the region of covariate values and quantile indices of interest:

\begin{example}
xsurf1 <- as.vector(piv.bsp.firstderiv$taus)
ysurf1 <- as.vector(piv.bsp.firstderiv$var.unique)
zsurf1 <- t(piv.bsp.firstderiv$point.est)
xsurf2 <- as.vector(piv.bsp.secondderiv$taus)
ysurf2 <- as.vector(piv.bsp.secondderiv$var.unique)
zsurf2 <- t(piv.bsp.secondderiv$point.est)
\end{example}

Finally, we create the three dimensional plots for:
$$
\left(\tau,w\right) \mapsto \partial^k g\left(\tau,w\right)/\partial w^k, \ \ \left(\tau,w\right) \in I,
$$
where $k \in \{1,2\}$, and $I$ is the region of interest. 

\begin{example}
par(mfrow = c(1, 2))
persp(xsurf1, ysurf1, zsurf1, xlab = "Quantile Index", ylab = "Age (months)",
	zlab = "Growth Rate", ticktype = "detailed", phi = 30, theta = 120, d = 5,
	col = "green", shade = 0.75, main = "Growth Rate (B-splines)")
persp(xsurf2, ysurf2, zsurf2, xlab = "Quantile Index", ylab = "Age (months)",
	zlab = "Growth Acceleration", ticktype = "detailed", phi = 30, theta = 120, 
	d = 5, col = "green", shade = 0.75, main = "Growth Acceleration (B-splines)")
\end{example}

\begin{figure}
\begin{center}
\includegraphics[width=\textwidth]{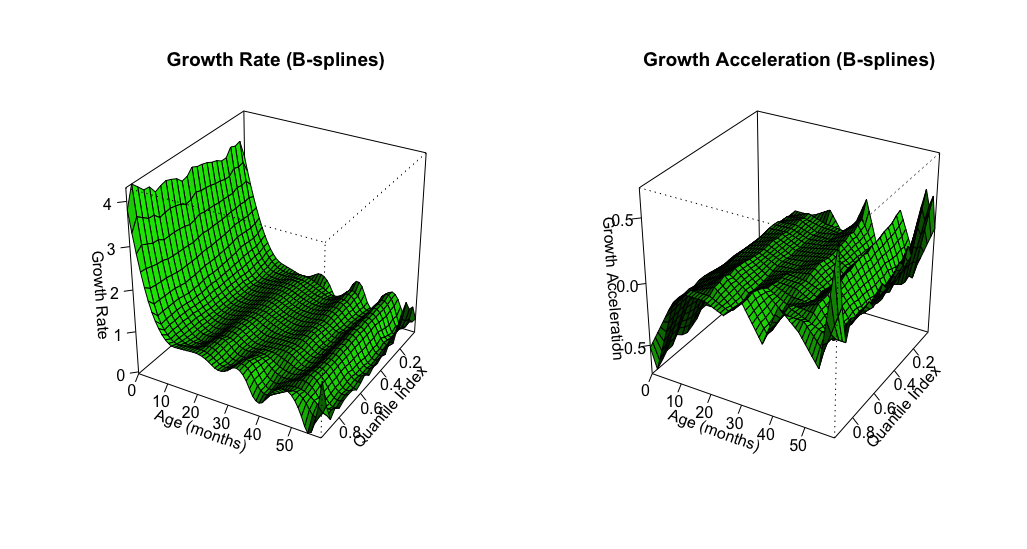}\caption{\label{fig:rate_acceleration}Growth Rate and Acceleration: estimates of the first and second derivatives of the conditional quantile function of height with respect to age.}
\end{center}
\end{figure}

These commands produce Figure \ref{fig:rate_acceleration}. Here, we see that the growth rate is positive at all ages and quantile indices. The growth rate decreases in the first few months of life and stabilizes afterwards, which can also be seen in the graph of growth acceleration. Growth acceleration is negative at young ages but stabilizes around zero at about 15 months. Both growth rate and growth acceleration are relatively homogeneous across quantiles at all ages.  Saved in \code{piv.bsp.firstderiv\$pvalues} and \code{piv.bsp.secondderiv\$pvalues} are the p-values from hypothesis tests to determine whether the first and second derivatives, respectively, are negative, positive, and equal to zero uniformly over the region of ages and quantile indices:

\begin{tabular}{cccc}
\code{Order of} &
\code{H0: Growth} &
\code{H0: Growth} &
\code{H0: Growth}\tabularnewline
\code{Derivative} &
\code{Rate <= 0} &
\code{Rate >= 0} &
\code{Rate = 0}\tabularnewline
\code{First Derivative} &
\code{0} &
\code{1} &
\code{0.042}\tabularnewline
\code{Second Derivative} &
\code{1} &
\code{0} &
\code{0.061}\tabularnewline
\end{tabular}

Thus, we reject at the 5\% level the null hypotheses that growth rate is negative, that growth rate is equal to zero, and that growth acceleration is positive
over all the first five years of the children's lives at all the quantiles of interest. We come close to rejecting at the 5\% level the null hypothesis that growth acceleration is equal to zero over all the first five years of the children's lives at all the quantiles of interest.

Similarly, we estimate the conditional quantile function over a region of covariate values and quantile indices, which corresponds to a growth chart in our application. Here, we use a fully saturated indicator basis for the series approximation to the nonparametric part of the model. We also compare the original estimates of the resulting growth chart to rearranged estimates that impose that the conditional quantile function of height is monotone in age and the quantile index. In this example, the conditional quantile function estimated using all data is nearly monotone without rearrangement.  To illustrate the power of rearrangement when estimates are not monotone, we use a subset of the data containing the first 1,000 observations:

\begin{example}
data.subset <- data[1:1000,]
detach(data)
attach(data.subset)
\end{example}

Now, we create the fully saturated indicator basis for \code{cage}:

\begin{example}
faccage <- factor(cage)
\end{example}

To perform estimation using this basis, we input \code{faccage} for \code{basis}:

\begin{example}
piv.fac.fun <- npqr(formula = form.par, basis = faccage, var = "cage", taus = taus,
	print.taus = print.taus, B = B, nderivs = 0, average = 0, alpha = alpha, 
	process = "none", rearrange = F, rearrange.vars="both", se = "conditional", 
	printOutput = F, method = "fn")
\end{example}

We also obtain the rearranged estimates with respect to age and the quantile index using the options of the command \code{npqr}. Note that we input \code{"both"} for \code{rearrange.vars}. This option performs rearrangement over quantile indices and age. Other allowable options are \code{"quantile"} (for monotonization over quantile indices only) and \code{"var"} (for monotonization over the variable of interest only). 

\begin{example}
piv.fac.fun.re <- update(piv.fac.fun, rearrange.vars = "both")
\end{example}

Now, we construct  three dimensional plots for the estimates of the conditional quantile function:
$$
\left(\tau,w\right) \mapsto Q_{Y \mid X}\left(\tau \mid x\right) = g\left(\tau,w\right) + v'\gamma\left(\tau\right), \ \ \left(\tau,w\right) \in I,
$$
where $v$ are evaluated at the sample mean for cardinal variables (\code{mbmi}, \code{breastfeeding}, \code{mage}, \code{medu}, and \code{edupartner}) and the sample mode for unordered factor variables (\code{faccsex}, \code{facctwin}, \code{faccbirthorder}, \code{facmunemployed}, \code{facmreligion}, \code{facmresidence}, \code{facwealth}, \code{facelectricity}, \code{facradio}, \\ \code{factelevision}, \code{facrefrigerator}, \code{facbicycle}, \code{facmotorcycle}, and \code{faccar}).

\begin{example}
xsurf <- as.vector(piv.fac.fun$taus)
ysurf <- as.vector(piv.fac.fun$var.unique)
zsurf.fac <- t(piv.fac.fun$point.est)
zsurf.fac.re <- t(piv.fac.fun.re$point.est)
par(mfrow = c(1, 2))
persp(xsurf, ysurf, zsurf.fac, xlab = "Quantile Index", ylab = "Age (months)",
	zlab = "Height", ticktype = "detailed", phi = 30, theta = 40, d = 5, 
	col = "green", shade = 0.75, main = "Growth Chart (Indicators)")
persp(xsurf, ysurf, zsurf.fac.re, xlab = "Quantile Index", ylab = "Age (months)",
	zlab = "Height", ticktype = "detailed", phi = 30, theta = 40, d = 5, 
	col = "green", shade = 0.75, main = "Growth Chart (Indicators, Rearranged)")
\end{example}

\begin{figure}
\begin{center}
\includegraphics[width = \textwidth]{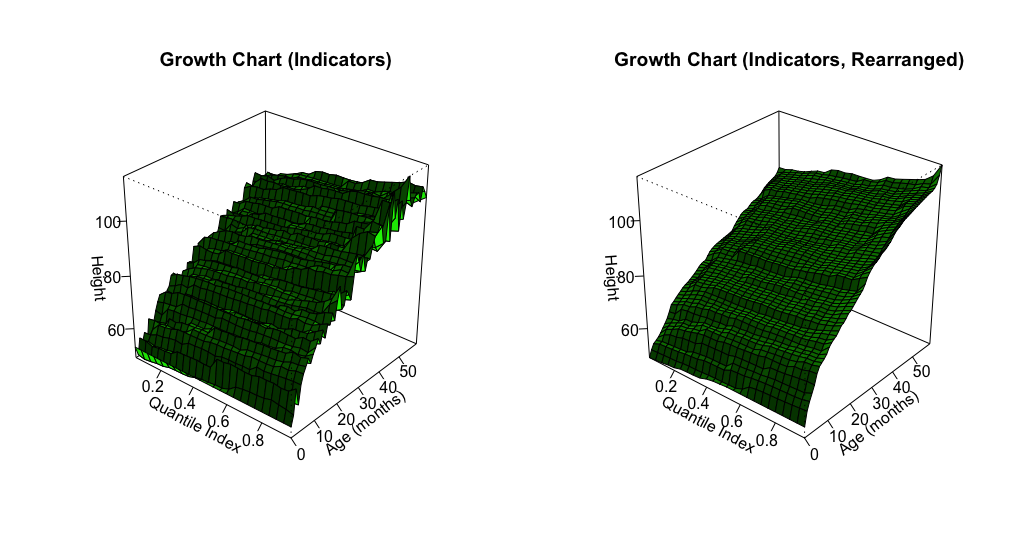}
\caption{\label{fig:growth_chart}Growth Chart, with and without Rearrangement: estimates of the conditional quantile function of height  based on a fully saturated indicator approximation with respect to age.}
\end{center}
\end{figure}

Figure \ref{fig:growth_chart} shows that the rearrangement fixes the non-monotonic areas of the original estimates.

\section{Conclusion}
In this paper we introduced the R package \pkg{quantreg.nonpar}, which implements the methods of \cite{bcf11} to estimate and make inference on partially linear quantile models. The package allows the user to obtain point estimates of the conditional quantile function and its derivatives based on a  nonparametric series QR approximation. Using pivotal,  gradient bootstrap,  Gaussian, and a weighted bootstrap methods, the user is also able to obtain pointwise and uniform confidence intervals. We apply the package to a dataset containing information on child malnutrition in India, illustrating the ability of \pkg{quantreg.nonpar} to generate point estimates and confidence intervals, as well as output that allows for easy visualization of the computed values. We also illustrate the ability of the package to monotonize estimates by the variable of interest and by quantile index.

\section{Acknowledgments}

We wish to thank Jim Ramsay for assistance with the \CRANpkg{fda} package (\cite{fda2014}), Roger
Koenker for sharing the data used in \cite{koenker2011}, and an anonymous referee for insightful comments.  We gratefully acknowledge research support from the NSF.

\bibliography{lipsitz-belloni-chernozhukov-fernandez-val}

\address{Michael Lipsitz\\
  Boston University\\
  Department of Economics\\
  USA\\}
\email{lipsitzm@bu.edu}

\address{Alexandre Belloni\\
  Duke University\\
  Fuqua School of Business\\
  USA\\}
\email{abn5@duke.edu}

\address{Victor Chernozhukov\\
  MIT\\
  Department of Economics\\
  USA\\}
\email{vchern@mit.edu}

\address{Iv\'an Fern\'andez-Val \\
  Boston University \\
  Department of Economics\\
  USA\\}
\email{ivanf@bu.edu}

\end{article}

\end{document}